\documentclass[aps,12pt,
		prd,
		reprint,
		twocolumn,
		superscriptaddress,
		shortbibliography,
		nofootinbib,
		floatfix,
		notitlepage
		]{revtex4-1}
\pdfoutput=1
\usepackage{amsmath,amssymb,amsfonts}

\usepackage{natbib}
\usepackage[T1]{fontenc}
\usepackage[latin1]{inputenc}

\usepackage[dvipsnames]{xcolor}

\usepackage{hyperref}
\hypersetup{colorlinks=true,citecolor=Cyan,urlcolor=Red}

\usepackage{mathrsfs}
\usepackage{bbm}
\usepackage{slashed}
\DeclareSymbolFont{rsfs}{U}{rsfs}{m}{n}
\DeclareSymbolFontAlphabet{\mathrsfs}{rsfs}
\usepackage[mathscr]{eucal}	
\usepackage[normalem]{ulem}
\usepackage{verbatim}
\usepackage{bm}

\usepackage{graphicx}

\newcommand{\be}{\begin{equation}}
\newcommand{\ee}{\end{equation}}
\newcommand{\bi}{\begin{itemize}}
\newcommand{\ei}{\end{itemize}}
\newcommand{\bea}{\begin{eqnarray}}
\newcommand{\eea}{\end{eqnarray}}
\newcommand{\ud}{\mathrm{d}}		

\newcommand{\LCm}{{\scriptscriptstyle -}} 
\newcommand{\LCp}{{\scriptscriptstyle +}}
\newcommand{\LCpm}{{\scriptscriptstyle \pm}}
\newcommand{\LCmp}{{\scriptscriptstyle \mp}}

\newcommand{\LCperp}{{\scriptscriptstyle \perp}}

\begin{document}
%
\title{Resummation of background-collinear corrections in strong field QED}

\author{James P. Edwards}
\email{jedwards@ifm.umich.mx}
\affiliation{Instituto de F\'isica y Matem\'aticas, 
Universidad Michoacana de San Nicol\'as de Hidalgo,
Edificio C-3, Apdo. Postal 2-82,
C.P. 58040, Morelia, Michoac\'an, M\'exico}

\author{Anton Ilderton}
\email{anton.ilderton@plymouth.ac.uk}
\affiliation{Centre for Mathematical Sciences, University of Plymouth, PL4 8AA, UK}

\begin{abstract}	
Scattering processes in laser backgrounds are degenerate to the emission of unobservable photons collinear with the laser. We identify processes and observables for which such degeneracies factorise and exponentiate, obtaining the leading-order intensity dependence of these inclusive observables at high laser intensity, correct to all orders in the fine-structure constant (all loops, all emissions). The results show an exponential intensity dependence distinct from that predicted by the Narozhny-Ritus conjecture on the high-intensity behaviour of quantum electrodynamics in strong fields. 
\end{abstract}
\maketitle

\section{Introduction}
%
It has been conjectured that {QED} perturbation theory breaks down in the presence of sufficiently strong background fields, even when these fields can be accounted for without approximation~\cite{Ritus1,Narozhnyi:1979at,Narozhnyi:1980dc,Fedotov:2016afw}.  This Narozhny-Ritus (`NR') conjecture arose in the context of intense laser-matter interactions, where it was observed that loop effects calculated in constant `crossed' fields (the simplest model of laser fields, obeying $E^2-B^2=E.B=0$, {and} denoted CCF), scaled dominantly not with powers of the fine structure constant $\alpha$, but with powers of $g:=\alpha\chi^{2/3}$ in which $\chi$ is, roughly, the product of field intensity and probe particle energy~\cite{Ritus1985a}: {for sufficiently high intensity $g$ can exceed unity, implying that all orders of perturbation theory must be included in any calculation. It has now been verified that certain $n$-loop self-energy contributions to the electron propagator in a CCF indeed scale as $g^n$~\cite{Mironov:2020gbi}.}

It is important to ask whether the NR conjecture applies in general backgrounds, or if it is a peculiarity of the (unphysical) CCF case. The `LCFA', an approximation which argues that any strong field may be approximated as locally constant and crossed~\cite{nikishov64,Ritus1985a}, suggests that the conjecture could hold generally: however, exactly solvable examples show that this cannot be the case~\cite{Ilderton:2019vot}. It is also known that the NR conjecture does not hold, outside of the CCF case, at high-energy~\cite{Podszus:2018hnz,Ilderton:2019kqp}.

{Making general statements about the conjecture is difficult, especially beyond CCF, if one cannot appeal to perturbation theory to any finite order. One may, though, try \textit{resumming} perturbative results. The state-of-the-art in a CCF is that the resummed  `bubble chain' of self-energy corrections to the electron propagator scales with $g\sqrt{\alpha}$ and $g^{3/2}\sqrt{\alpha}$~\cite{Mironov:2020gbi}. We note, for what follows, that this does not appear to be an exponentiation of the perturbative series.}

{We will here investigate high-intensity behaviour and resummation in fields more general that a CCF. Our investigation is based on degenerate processes and inclusive observables. As motivation, recall that in QED any process is degenerate with that in which, in addition, arbitrarily many soft, undetectable, photons are emitted}. These degeneracies are summed over to obtain \textit{inclusive} observables. Consistency {then} requires that soft contributions from photon loops also be included; doing so yields infra-red (`IR') finite, inclusive observables~\cite{Yennie:1961ad,Weinberg:1965nx}. {When a background laser is present, however, there are additional degeneracies: in a given scattering process, any photons emitted almost collinear with an intense laser and with energies not dissimilar from it, will be indistinguishable from laser photons, or masked by the high flux of the field. Physical observables should be made inclusive with respect to these emissions  -- we will account for them here.}
	
The immediate difficulty to confront is that there is no exact method for calculating scattering amplitudes in general \textit{high-intensity} fields. (As indicated above, relying on the LCFA conflates the validity of the NR conjecture with the limited accuracy of that approximation~\cite{Harvey:2014qla,DiPiazza:2017raw,Ilderton:2018nws}.) Consider then plane waves, in which amplitudes can be calculated exactly at arbitrary intensity. While lacking in realistic spatial structure (focussing), plane waves have a direction and typical frequency, so that it makes sense to speak of degeneracies due to photon emission \textit{collinear} with {the plane wave}. By summing over such degeneracies we will {here} calculate {inclusive observables in plane wave backgrounds to all orders in $\alpha$. Crucially, we will show that these observables introduce a nontrivial intensity dependence to which the CCF and LCFA are blind, which leads to a high-intensity scaling distinct from that predicted by the NR conjecture.}

{The delicate part of our calculation is the identification of relevant scales and approximations: this and final results are presented in Sect.~\ref{SECT:RESULTS}. The calculation itself uses now-standard literature methods~\cite{Ritus1985a,Seipt:2017ckc}, and is not dissimilar to textbook calculations of soft IR effects (though the physics is different -- see~\cite{Laenen:2020nrt} and references therein for recent work on soft and soft-collinear factorisation in QED).  As such, details are given in the appendices. The calculation, being all-orders in $\alpha$, is still challenging, so we keep track throughout only of the \textit{leading} intensity dependence. We conclude in Sect.~\ref{SECT:CONCS}.}

\section{Results}\label{SECT:RESULTS}

\paragraph*{Laser-collinear degeneracy.}
Consider a collision between particles and a laser pulse, the latter propagating in the direction $n_\mu=(1,0,0,1)$. An angular cutoff $\theta_0$ about this direction may be defined by e.g.~the laser opening angle: any photons emitted within this cone will be indistinguishable from laser photons if their energy is similar to that of the laser, typified by $\omega_0$ (order 1~eV for optical beams), {due to either limited detector resolution in energy and angle, or to over-saturation from the large flux of the intense field. We assume that any lower energy photons emitted within the same cone are also unobservable -- Notch filters allow for more sophisticated setups~\cite{Notch1,Notch2}, but we consider only the simplest here.} 

A photon of momentum $q_\mu$ emitted within an angle {$\theta<\theta_0$} relative to the laser direction obeys, for $\theta_0$ small,  $q_\mu = q_0 n_\mu + \mathcal{O}(\theta_0)$, or, in explicit components, $q_\LCm:=(q_0-q_3)/2 \simeq 0$, $q_\LCperp :=(q_1,q_2) \simeq 0$, $q_\LCp :=(q_0+q_3)/2 \simeq q_0$. The leading contribution of such photons in scattering amplitudes may thus be found by setting $q_\mu = q_\LCp n_\mu$ in calculations. {Another way to see this is to observe that these photons obey $|q_\LCperp|\simeq q_\LCp \theta$ and $q_\LCm \simeq \frac{\theta^2}{4}q_\LCp$, which yields, given that $\theta$ is itself restricted to small values, the hierarchy ${q_\LCm < q_\LCperp < q_\LCp}$. For each such photon probabilities/cross-sections are integrated over, using the usual on-shell measure,}
\be\label{coll-int}
	 \int\!\frac{\ud^3{\bf q}}{(2\pi)^32q_0} \to \frac{\theta_0^2}{4(2\pi)^2}\int\limits_0^{\omega_0}\! \ud q_\LCp q_\LCp \;.
\ee
\begin{figure}[t!]
\includegraphics[width=\columnwidth]{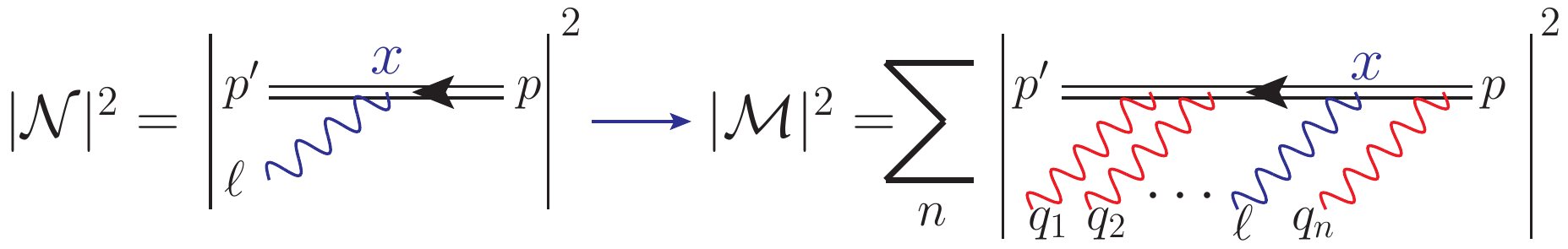}
\caption{\label{FIG:NLC} Exclusive (left) and inclusive (right) nonlinear Compton scattering at tree level. The double-lines indicate Volkov electron wavefunctions, dressed exactly and to all orders by the background~\cite{volkov35}. Photons with momentum $q_j^\mu$ are degenerate with the laser.}
\end{figure}
\noindent\paragraph*{Laser-collinear emissions.}
We consider nonlinear Compton scattering (NLC)~\cite{nikishov64,Kibble:1965zza,Ritus1985a,Harvey:2009ry,Boca:2009zz,Seipt:2010ya,Mackenroth:2010jr,Seipt:2017ckc}, that is the emission of a photon, momentum $\ell_\mu$, from an electron, momentum $p_\mu$, traversing a plane wave described by the two-component transverse potential $\mathsf{a}(n\cdot x)$ (the integral of the electric field~\cite{Dinu:2012tj}). This \textit{exclusive} process shows, for constant crossed fields, the high-intensity scaling associated with the NR conjecture. Our interest is in the corresponding \textit{inclusive} process which accounts for the additional emission of arbitrarily many laser-collinear degenerate photons described above; see Fig.~\ref{FIG:NLC}.  In the corresponding scattering amplitudes, there are nontrivial integrals over `lightfront time' $x_j^\LCp:=n \cdot x_j$ at each emission vertex $x_j^\mu$, due to the spacetime dependence of the background {plane wave}. (See Appendix~\ref{APPA} for details.) In particular, the degenerate emissions introduce a lightfront time-ordered dependence on $x_j^\mu$ which prevents their contributions from factorising out in general, as would be the case with leading soft factors.

\begin{figure}[b!]
	\includegraphics[width=0.9\columnwidth]{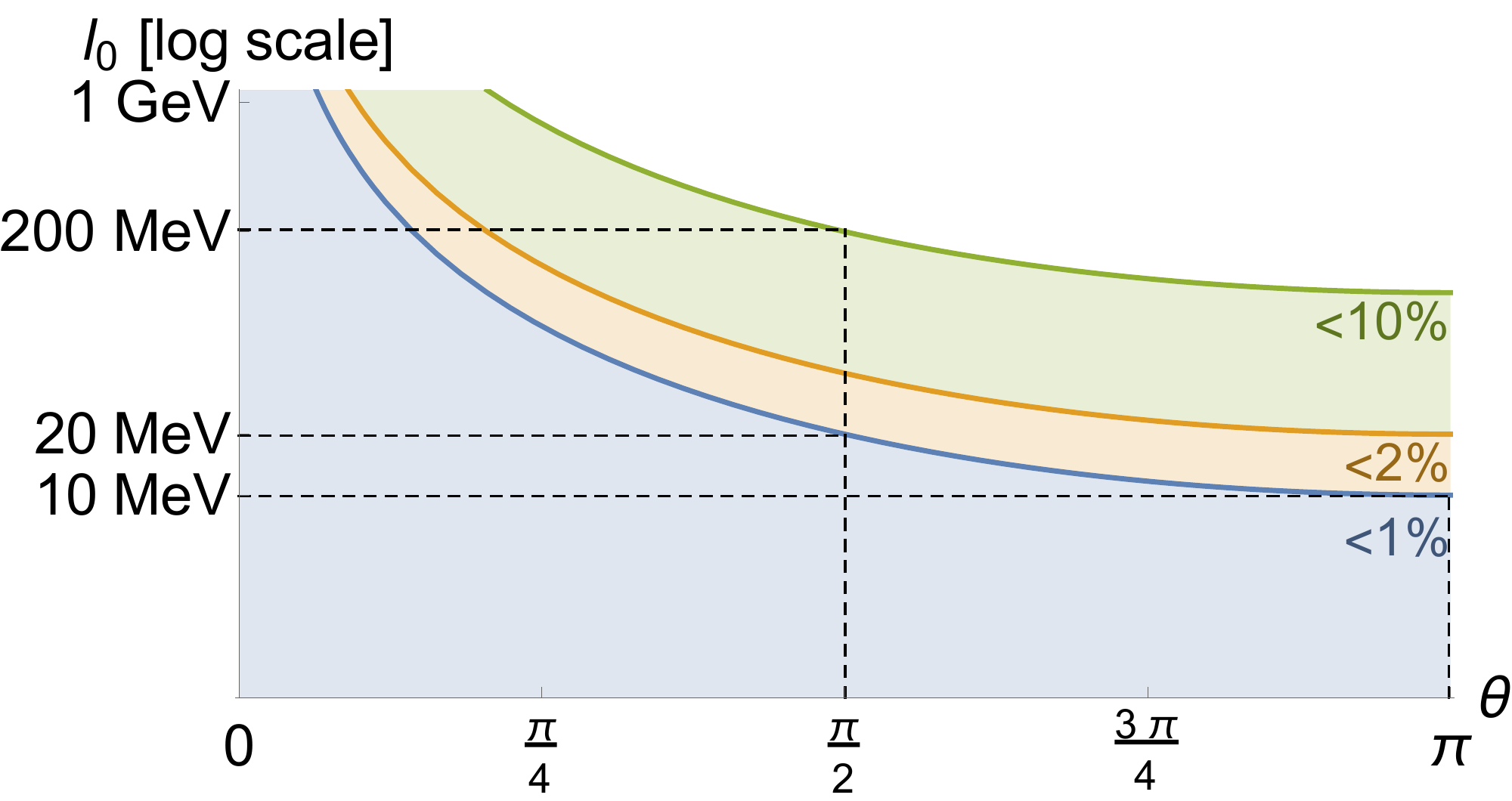}
	\caption{\label{Fig:approx} The error induced by our approximation, in terms of the observed photon energy $\ell_0$ and emission angle $\theta$ (relative to the laser direction), in the case of a head-on collision between a $1$~GeV electron and a laser pulse. Each curve bounds a region of percentage error, as highlighted.}
\end{figure}
It is however possible to find observables for which collinear corrections do factorise. In that part of the observable photon spectrum for which $\ell_\mu$ obeys ${s \equiv n\cdot \ell / n \cdot p \ll 1}$, we find that the effect of adding a collinear emission is simply to multiply the amplitude by a scalar factor, up to an error of order $s$. Let us analyse this restriction. It is important to emphasise that $\ell_\mu$ is supposed to be observable, and small $s$ \emph{does not} mean that $\ell_\mu$ is itself degenerate with the laser. To illustrate this, suppose we insist on $s \leq 0.01$, which means a $1\%$ error induced by our approximation. Then for a head-on collision of a 1~GeV electron with the laser,  see Fig.~\ref{Fig:approx},  we are restricted to considering emitted photon energies of $<10$~MeV when {those photons are} scattered \textit{forward} of the electron, and $<20$~MeV for scattering at right-angles to the collision axis: these emissions are not laser-degenerate, and are experimentally measurable. 

It is also important to stress that, in considering observable photons at small $s$, and degenerate emissions which have $s =0$, we are focussing on that part of the emission spectrum in which the LCFA is known to \text{fail}~\cite{Harvey:2014qla,DiPiazza:2017raw,Ilderton:2018nws,Heinzl:2020ynb}.  {To be explicit, let $\mathcal{N}$ be the {tree-level} NLC amplitude, and let $a_0 \sim |{\sf a}|/m$ be the `dimensionless intensity parameter' characterising the strength of the background. The LCFA predicts that the small-$s$ emission probability in NLC behaves as~\cite{nikishov64,Ritus1985a}
\be\label{s1}
	\frac{\ud \mathbb{P}}{\ud s}\bigg|_{s\ll 1} \overset{!}{\sim} \int\!\ud^2p'_\LCperp |\mathcal{N}|^2_\text{LCFA} \sim \frac{a_0^{2/3}}{s^{2/3}} \;,
\ee
at high intensity, exhibiting the typical 2/3-power scaling of the NR conjecture.  However, it can be shown without approximation that the true behaviour is~\cite{DiPiazza:2017raw}
\be\label{s2}
	\frac{\ud \mathbb{P}}{\ud s}\bigg|_{s\ll 1} \sim \int\!\ud^2p'_\LCperp |\mathcal{N}|^2 \sim a_0^2 \;.
\ee
As such, the high-intensity behaviour of our inclusive observables may be expected to, and indeed will, differ from that in constant fields.}

{Returning to our calculation, laser-collinear emissions both factorise and exponentiate in the considered part of the spectrum.} We write a prime on amplitudes to indicate that we restrict to this regime. Let $|\mathcal{M}|^2$ {be the} inclusive sum over (mod squared) amplitudes with {any number} of collinear emissions, all at tree level, see Fig.~\ref{FIG:NLC}. We show in Appendix~\ref{APPA} that the leading \textit{intensity} dependence of $|\mathcal{M}'|^2$ is, for $\tilde{\sf a}$ the Fourier transform of ${\sf a}$,
\be\label{real}
	|\mathcal{M}'|^2 \sim \exp \bigg[  \frac{\alpha \theta_0^2}{4\pi}\int\limits_0^{\omega_0}\! \ud q_\LCp\,  q_\LCp \frac{ |\tilde{\sf a}(q_\LCp)|^2}{(n\cdot p)^2} \bigg] |\mathcal{N}'|^2 \;.
\ee
The exponent is positive, and we conclude that degenerate emissions \textit{enhance} the measured emission probability (as soft emissions do~\cite{Yennie:1961ad,Weinberg:1965nx}). This enhancement is exponentially increasing with intensity -- by unitarity, this must be compensated for by a similar exponential factor from loop corrections. We confirm this below, but (\ref{real}) already telegraphs the final result -- the high-intensity scaling of inclusive observables will be exponential.

\begin{figure}[b!!]
\includegraphics[width=0.9\columnwidth]{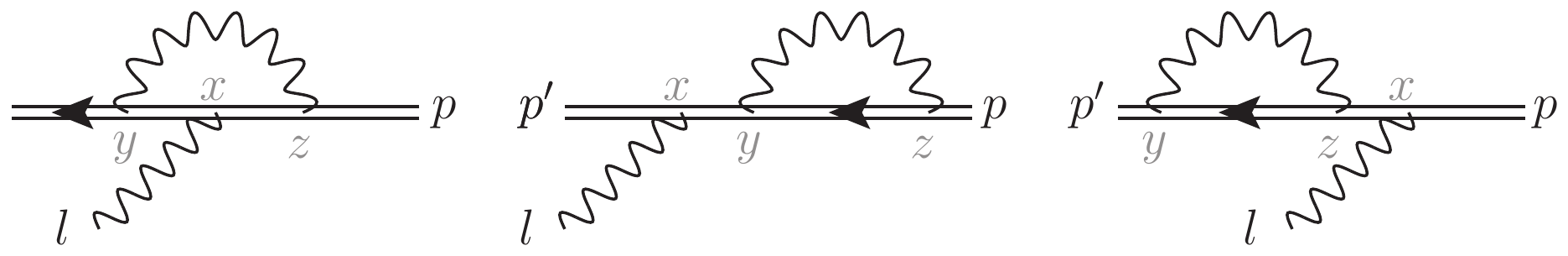}
\caption{\label{FIG:LOOPS} The 1-loop contributions to nonlinear Compton scattering; the vertex correction ({left} diagram) and self-energies.}\end{figure}

\paragraph*{Loop corrections.}
%
%
%
As (\ref{real}) contains all orders of $\alpha$, we must for consistency also consider all orders of loop corrections to NLC; to illustrate, the relevant one-loop diagrams are shown in Fig.~\ref{FIG:LOOPS}; two self-energies~\cite{Ritus1985a,Meuren:2015lbz} and the vertex correction~\cite{Morozov:1981pw,Gusynin1999,DiPiazza:2020kze}.  Note that we are only interested in $a$-dependent terms, which are UV-finite~\cite{Naroz1968,Becker}. {We cannot, and do not aim to, calculate exact loop effects to all orders in $\alpha$, but only to include effects from \textit{laser-collinear} virtual photons which i) must compete with the real emissions investigated above and ii) will be incorrectly captured by the LCFA, as real emissions are.}

{We begin as we did for real emissions by addressing the momentum integration region for virtual laser-collinear photons, which is again more involved than for soft corrections. We need to identify the full region over which the approximations we made for real emissions continue to hold. This is simplified, as described in Appendix~\ref{APPB}, by performing the contour integral in the photon propagator, which allows us to take the virtual photons to obey the on-shell relation $q_\LCp = q_\LCperp^2/4q_\LCm$.} 
	{Collinearity requires that the transverse momentum $q_\LCperp$ {be less} than some \textit{absolute} cutoff, as otherwise the photon momentum will no longer be approximately degenerate with the laser (and distinct from the observable momenta in the game).}
	It can be checked that, for the \text{real} emissions above, the energetic restriction on the photon momentum implies $q_\LCm < \omega_0\theta_0^2/4$ and $|q_\LCperp| < \omega_0 \theta_0$; {but these inequalities, and the hierarchy $q_\LCm < q_\LCperp < q_\LCp$, are obeyed over the larger region of momentum space shown in Fig.~\ref{FIG:options}, which encloses the real emission region, but also admits increasingly high photon energy at smaller (closer to laser-collinear) emission angles.} (Essentially, we can make the virtual photon momentum $q_\LCm$ small by taking $q_\LCp$ to be arbitrarily large\footnote{{In contrast to soft effects, we can integrate over arbitrarily high energies, but we repeat that the terms of interest are UV finite (they simply yield Fourier transforms).}}, provided $q_\LCperp$ is bounded.) The complete region over which we may integrate virtual photon momenta while staying within our approximation is then conveniently characterised by $|q_\LCperp| \simeq q_0 \theta < \omega_0 \theta_0$. We note that this includes a region with $\theta>\theta_0$ but $q_0<\omega_0$ ({not marked} in the Figure), which takes us away from collinear and into the IR; we neglect this region, as we are not here keeping track of IR effects (but see the appendices.)
 
{We find, as at tree level (see Appendix~\ref{APPB} for the full calculation), that these laser-collinear virtual photons} introduce lightfront-time ordered terms into the \textit{integrand} of the NLC amplitude, which prevents factorisation in general. Making the same assumption as above, i.e.~restricting attention to that part of the emission spectrum with $n\cdot l \ll n\cdot p$, we again obtain factorisation. 

Integrating over the region marked in Fig.~\ref{FIG:options}, we find that the contribution from the virtual photons entirely cancels that from real emissions. {(Recall that in the IR}, soft divergences cancel between emissions and loops~\cite{Yennie:1961ad,Weinberg:1965nx}.) What remains yields our final result for the all-loops inclusive amplitude:

\begin{figure}[t!]
\includegraphics[width=0.8\columnwidth]{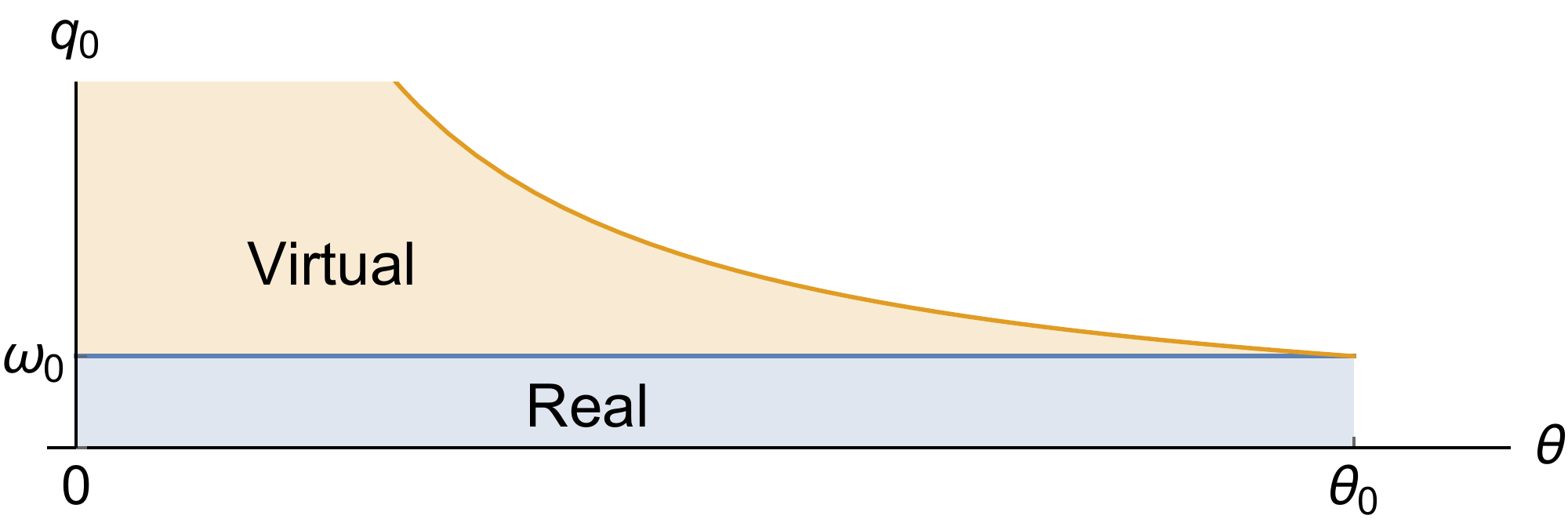}
\caption{\label{FIG:options} Region of momentum integration for virtual laser-collinear photons; this covers both the real emission region (blue) and photons and a region of high energy, but small emission angle (orange).}
\end{figure}

\be\label{everything2}
 	|\mathcal{M}'|^2_\text{all loops} \sim \exp \bigg[  -\frac{\alpha \omega^2_0\theta_0^2}{4\pi}\int\limits^{\infty}_{\omega_0}\! \frac{\ud q_\LCp}{q_\LCp} \frac{ |\tilde{\sf a}(q_\LCp)|^2}{(n\cdot p)^2}  \bigg]  |\mathcal{N}'|^2 \;.
\ee
We have thus found that inclusive corrections lead to an \textit{exponential} intensity dependence at high intensity.  Let us compare with~\cite{Mironov:2020gbi} in {which one-loop polarisation operator insertions to the electron propagator (the bubble chain) are resummed exactly in a CCF. This gives essentially the elastic scattering amplitude, whose imaginary part is the probability of NLC}.  The resummed amplitude still scales with a positive, fractional power of intensity: curiously, {this might suggest} that resummed results, and observables constructed from them, {could} still be made arbitrarily large at high intensity. In our resummed result, on the other hand, the exponential damping controls the high-intensity behaviour, in-line with both unitarity and the behaviour of resummed soft corrections.  {See Appendix~\ref{APP:PHEN} for the phenomenology of the corrections implied by (\ref{everything2}); for upcoming experiments, these are certainly negligible, but this is not the regime of interest.}

{Our results are not in contradiction to CCF results.  As in (\ref{s1})--(\ref{s2}), the small-$s$ behaviour in genuinely constant fields is very different to that in any field which switches on and off asymptotically, {as is also true of soft effects~\cite{Dinu:2012tj}}. Indeed the small-$s$ region we consider is precisely that in which the locally constant field approximation breaks down~\cite{Harvey:2014qla,DiPiazza:2017raw,Ilderton:2018nws}.   Our results show that collinear $n$-loop (and emission) effects beyond CCFs scale with powers not of $\alpha^n $ but with $(\alpha  a_0^2)^n$, which grow faster with intensity than the $\alpha a_0^{2/3}$ scaling predicted by CCF results. We have shown that these effects \textit{exponentiate} upon resummation, changing for example the exclusive NLC intensity dependence (\ref{s2}) to the inclusive result
	\be
	\frac{\ud \mathbb{P}_\text{inc}}{\ud s}\bigg|_{s\ll 1} \sim \int\!\ud^2p'_\LCperp |\mathcal{M'}|^2 \sim 		a_0^2 \exp(-\alpha a_0^2 \cdot \text{const})\;.
	\ee
}
\section{Discussion}\label{SECT:CONCS}

{In the context of the NR conjecture, on the high-intensity behaviour of QED in external fields, we have investigated inclusive observables in the nonlinear Compton scattering of a photon from an electron in an intense laser pulse, in which we include laser-collinear (degenerate) emissions.
	
	To assess the relevance of the conjecture beyond the case of constant crossed fields (CCFs), we have specifically addressed that part of the photon emission spectrum in which locally constant field approximations, commonly used to generalise CCF results to general fields, are well-known to break down~\cite{Harvey:2014qla,DiPiazza:2017raw,Ilderton:2018nws}. In this part of the spectrum, laser-collinear emissions were found to factorise and, resummed to all-orders in $\alpha$ (all emissions, all loops), yield an \textit{exponential} dependence on intensity (and not on the product $\chi$ of energy and intensity) which is distinct from any behaviour seen to date for constant fields, even when resummed, and cannot be captured by locally constant field approximations.}

{As such our results do not contradict CCF results. Rather we have shown that high-intensity behaviour, outside of a CCF, can differ significantly from that predicted by the NR conjecture, both at finite order in perturbation theory and after resummation. Our findings thus reinforce the idea that the conjecture has limited relevance beyond the CCF case~\cite{Podszus:2018hnz,Ilderton:2019kqp,Ilderton:2019vot}. (While our calculation of degenerate processes assumed a pulsed plane wave background, see also Appendix~\ref{APP:PHEN}, similar, in fact richer, degeneracies clearly exist for more realistic fields.)}

{We can view our results from a different angle. As part of our calculation we have shown that, outside of CCFs, higher {loops} in strong backgrounds can induce effects scaling with powers not of $\alpha$, but with $\alpha a_0^2$, which is a stronger scaling than that predicted by the NR conjecture. This also suggests a breakdown of perturbation theory. We have shown that the considered effects can be resummed, upon which they exponentiate, and bring with them both a nontrivial intensity dependence due to the presence of the background, {and} a portion of the usual soft IR divergence. This can be cancelled by also including laser-collinear degenerate emissions, which also introduce an extra intensity dependence; what remains is of course (\ref{everything2}) with its exponential high-intensity behaviour. Thus, outside of constant fields, resummation yields well-behaved results consistent with unitarity and in-line with the behaviour of soft corrections.}

 {We restricted to a particular part of the NLC spectrum. It could well be that in {another part} of the spectrum (e.g.~high lightfront energy, rather than low) the resummed high-intensity behaviour is different again. It would be interesting to investigate this. The high-intensity behaviour of other processes and more general observables, and their degeneracies, remain an open question.  {Let us briefly discuss some effects which should be included in future calculations.
 
In general, there are other sources of (potentially collinear) radiation, such as vacuum emission~\cite{Karbstein:2014fva,Gies:2017ygp}, but in a plane wave background such ``dressed tadpole'' contributions are vanishing~\cite{Ahmadiniaz:2019nhk}, hence there is nothing to include in our case. One might expect that such emissions would, in any case, be less significant than those from charged particles being rapidly accelerated by the strong field, and indeed an abundance of produced photons has been identified as an experimental signature of e.g.~Schwinger pair production~\cite{Otto:2016xpn,Aleksandrov:2019irn}. A preliminary estimate of the impact of soft emissions could be obtained by evaluating the Weinberg soft factors~\cite{Weinberg:1995mt} for our processes, following the methods in~\cite{Ilderton:2012qe}, and seeing how $a_0$ enters. It may be that more comprehensive results could be obtained in constant crossed fields using the resummation techniques of~\cite{Mironov:2020gbi}; however, the crossed field approximation is well-known to fail in the IR.
 
It is expected that back-reaction on the driving laser field becomes significant for scattering in the high-intensity regime~\cite{Seipt:2016fyu}. How back-reaction impacts the NR conjecture remains a challenging open question. We note that no analogous perturbative breakdown appears in the simplest toy models for which back-reaction can be included~\cite{Ekman:2020vsc}, but this may be due to the simplicity of the model. }

We can draw one final conclusion. Experimental proposals~\cite{Yakimenko:2018kih,Baumann:2018ovl,Blackburn:2018tsn,DiPiazza:2019vwb} for observing signals of the (anticipated) non-perturbative physics at $\alpha\chi^{2/3}\gtrsim 1$ often rest on observing `some' deviation from literature predictions, as intensity is made large. These predictions are, though, based on \textit{lowest-order} perturbative calculations of \textit{exclusive} observables (or phenomenological models built from them), while any measured observable will be partially \textit{inclusive}, since no experiment has 100\% detector coverage. Our results show that it is not enough, in making experimental predictions at high intensity, to take deviations from lowest order predictions as evidence for entering the non-perturbative regime:  there are other effects with a competing intensity dependence.}

\begin{acknowledgments}
\noindent\textit{A.I.~thanks Tom Heinzl and Stuart Mangles for useful discussions. J.P.E. acknowledges funding from a CIC-U.M.S.N.H. project.  A.I.~is supported by The Leverhulme Trust, project grant RPG-2019-148. Both authors received support from the Royal Society through Newton Mobility Grant NMG\textbackslash R1\textbackslash 180368}.
\end{acknowledgments}


%


\clearpage \onecolumngrid
\appendix
\label{SECT:APPENDIX}

\section{laser-collinear emissions and degeneracy}\label{APPA}
\renewcommand{\thesubsubsection}{\roman{subsubsection}}

\subsubsection{Notation, conventions, and the background field}
Our focus is on degeneracies due to unobservable photon emissions that are collinear with (the photons in) a background plane wave field, of arbitrary temporal profile, but compact support; a pulse. The field depends on $n\cdot x$ where $n_\mu$ is a null vector, $n^2=0$. We can always choose $n\cdot x = x^0+x^3$, lightfront time. We use lightfront coordinates $x^\LCpm = x^0 \pm x^3$, $x^\LCperp = (x^1,x^2)$ for position and $p_\LCpm=(p_0\pm p_3)/2$, $p_\LCperp=(p_1,p_2)$ for momenta. As such $v^\LCpm=2v_\LCmp$ for all vectors $v^\mu$. The background is $a_\mu(x) = \delta_\mu^\LCperp {\sf a}_\LCperp(x^\LCp)$ in which the two transverse components~${\sf a}_\LCperp$ arise as lightfront time integrals of the two electric field components of the plane wave, with $a_\mu(-\infty)=0$. For a typical (e.g.~ Gaussian) pulse shape, the field will be characterised by some dimensionless peak field strength $a_0$, i.e.~${\sf a} (x^\LCp) \sim m a_0$, which is the effective coupling between the background and fermions. $a_0$ easily exceeds unity in modern laser pulses.

\subsubsection{Classical and quantum particle dynamics}
The on-shell kinetic momentum of a classical particle in the plane wave (the \textit{exact} solution to the Lorentz force equation) is, for $p_\mu$ {the initial} momentum before entering the wave,   
\be\label{Lorentz}
	\pi_\mu(x^\LCp) = p_\mu - a_\mu(x^\LCp) + n_\mu \frac{2p\cdot a(x^\LCp) -a(x^\LCp)^2}{2n\cdot p} \;.
\ee
In the quantum theory, scattering amplitudes are calculated in the Furry picture, which treats the coupling $a_0$ as part of the `free' theory, i.e.~without approximation, and then treats the coupling~$e$ between dynamical (quantised) fields in perturbation theory, as normal. (It is this perturbative expansion which is conjectured to break down at sufficiently high field strength, at least in constant crossed fields.) Hence the interaction vertex is $-ie \gamma^\mu$ as usual, but the fermion propagator and external legs become dressed by the background. (For a recent overview see~\cite{Seipt:2017ckc}.)   The asymptotic wavefunction describing an incoming electron of initial momentum $p_\mu$ in amplitudes is
\be\label{pi-p}
	\psi_p(x) = \bigg[1 + \frac{\slashed{a}(x^\LCp)\slashed{n}}{2n\cdot p}\bigg] u_p\exp\bigg(-ip\cdot x-i\int\limits^{x^\LCp}_{-\infty} \frac{2p \cdot a-a^2}{2n \cdot p}\bigg) \;,
\ee
{where $u_p$ is the usual u-spinor}. For discussions of the physical content of these wavefunctions see for example~\cite{Dinu:2012tj,Seipt:2017ckc,Ilderton:2020gno}. The corresponding fermion propagator is
\be
	S(x,y) = i \int\!\frac{\ud^4 p}{(2\pi)^4} \bigg[1 + \frac{\slashed{n}\slashed{a}(x^\LCp)}{2n\cdot p}\bigg] \frac{ \slashed{p}+m }{p^2-m^2+i\epsilon}\bigg[1 + \frac{\slashed{a}(y^\LCp)\slashed{n}}{2n\cdot p}\bigg]e^{-ip\cdot(x-y)-i\int\limits^{x^+}_{y^\LCp} \frac{2p.a-a^2}{2n.p}} \;.
\ee

\subsubsection{{Laser-collinear emissions}}
{We recall that the Lorentz-invariant measure for on-shell photons may be written, in terms of cartesian coordinates or lightfront coordinates, as 
\be\label{os-mott}
{	\ud q_\text{o.s.} := \frac{\ud^3{\bf q}}{(2\pi)^32q_0}}  = \frac{\ud^2q_\LCperp\ud q_\LCpm}{(2\pi)^32q_\LCpm} \Theta(q_\LCpm) \;,
\ee
with the un-integrated momentum component being determined by the mass-shell condition $q_\mu q^\mu = 0$. In the text and in what follows we will use each of the three forms of the on-shell measure above, as necessary. 

Since the plane wave is univariate, depending on only $n\cdot x= x^\LCp$, the only nonzero laser momentum component is $q_\LCp$. As such laser-degenerate photons must have $\{q_\LCm,q_\LCperp\} \simeq 0$. How these conditions are imposed, and to what tolerance, depends on the situation of interest. As in the main text, real photons degenerate with the laser obey $q_0\leq \omega_0$ and $\theta<\theta_0$ for $\omega_0$ and $\theta_0$ given cutoffs in energy and emission angle relative to the laser direction, respectively. Using cartesian coordinates, this gives a measure
\be\label{mott1}
	\int\!\frac{\ud^3{\bf q}}{(2\pi)^32q_0} \to \frac{1}{2(2\pi)^3}\int\limits_0^{\omega_0} \ud q_0 \, q_0 \int\limits_0^{\theta_0}\ud \theta \sin\theta\int\limits_0^{2\pi}\!\ud\phi \to \frac{\theta_0^2}{16\pi^2}\int\limits_0^{\omega_0} \ud q_0 \, q_0 \;,
\ee
where, in the final step, we have assumed $\theta_0\ll1$ and truncated the integrand, whatever it may be, to zeroth order in $\theta_0$. In lightfront coordinates, which are often more useful in our investigation, the energetic and angular restrictions above reduce to, again for $\theta_0$ small, $q_\LCp<\omega_0$ and $|q_\LCperp| < q_\LCp \theta_0$. Hence our integral may also be written
\be\label{mott2}
	\int\! \frac{\ud^2q_\LCperp\ud q_\LCp}{(2\pi)^32 q_\LCp } \Theta(q_\LCp) \to \frac{1}{2(2\pi)^3}\int\limits_0^{\omega_0}\!\frac{\ud q_\LCp}{q_\LCp} \int\!\ud^2 q_\LCperp \Theta(q_\LCperp - q_\LCp \theta_0)  \to \frac{\theta_0^2}{16 \pi^2}\int\limits_0^{\omega_0}\!\ud q_\LCp\, q_\LCp \;,
\ee
which agrees with (\ref{mott1}) because $q_0=q_\LCp$ to leading order in $\theta_0$}.

\subsubsection{{Laser-collinear emission amplitudes}}
Consider an incoming electron leg of momentum $p_\mu$ in a given scattering amplitude, connecting to a vertex at position $x^\mu$, and so represented by a Volkov wavefunction $\psi_p(x)$. On that leg we add a laser-collinear emission. In the amplitude, $\psi_p(x)$ is then replaced by (stripping off the photon polarisation vector)
\be\label{FromLeft}
	\int\!\ud^4 y\, S(x,y) \gamma^\mu e^{i q\cdot y} \psi_p(y)  = e^{i q \cdot x}\bigg[ \frac{i \slashed{n} \gamma^\mu}{2n\cdot p} + \int\limits_{-\infty}^{x^\LCp}\!\ud y^\LCp\, \frac{\pi^\mu(y)}{n\cdot p} e^{-iq_\LCp(x-y)^\LCp}\bigg] \psi_p(x) \;,
\ee
in which the left hand expression follows from the Feynman rules, and the right hand side is \textit{exact} for $q_\mu \equiv q_\LCp n_\mu$, collinear with the plane wave. If $q_\mu$ deviates from collinear in some small parameter, as in Appendix~\ref{APPA}\textit{iii}~above,   then the right hand side of (\ref{FromLeft}) is correct to leading order in that parameter. For emission from an outgoing leg of momentum $p'_\mu$ we have, similarly,
\be\label{FromRight}
	\bar{\psi}_{p'}(x) \to \int\!\ud^4 y\, \bar\psi_{p'}(y) \gamma^\mu e^{i q\cdot y}S(y,x) =
	 \bar\psi_{p'}(x)  \bigg[ \frac{i \gamma^\mu \slashed{n} }{2n\cdot p'} + \int\limits^{\infty}_{x^\LCp}\!\ud y^\LCp\, \frac{\pi^{\prime \mu}(y)}{n\cdot p'} e^{-iq_\LCp(x-y)^\LCp}\bigg] e^{i q \cdot x} \;,
\ee
in which $\pi'$ depends on $p'$ as $\pi$ depends on $p$ in (\ref{pi-p}). We remark that damping factors are understood to be in place, in these expressions, for the terms in $\pi^\mu$ which are field-independent~\cite{Boca:2009zz,Ilderton:2020rgk}. These are the terms contributing to the vacuum limit, where the factor in square brackets of (\ref{FromLeft}) reduces to
\be\label{vaclimit}
	i \frac{\slashed{n}\gamma^\mu}{2n\cdot p} - i\frac{p^\mu}{p\cdot q - i \epsilon}  \;,
\ee
and similarly for (\ref{FromRight}), which we recognise from textbook discussions of collinear emission factors, see e.g.~\cite{Schwartz:2013pla}. Note that the second term in (\ref{vaclimit}) carries a factor of $1/q_\LCp$, equivalent to $1/q_0$ here, which is where the usual IR divergence arises\footnote{While there are, strictly, no \textit{collinear} divergences in massive QED, we mention that, contrary to what is often inferred from the Lee-Nauenberg theorem, the cancellation of collinear divergences in massless theories is far from resolved, {see~\cite{Lavelle:2005bt,Lavelle:2010hq} and also below}.}. We already know that such divergences drop out of inclusive observables, but we will keep track of them for the time being. The two results (\ref{FromLeft}) and (\ref{FromRight}) are the basic building blocks of the following calculations.

\subsubsection{Laser-collinear emission correction to nonlinear Compton}
\begin{figure}[t!]
\includegraphics[width=0.45\textwidth]{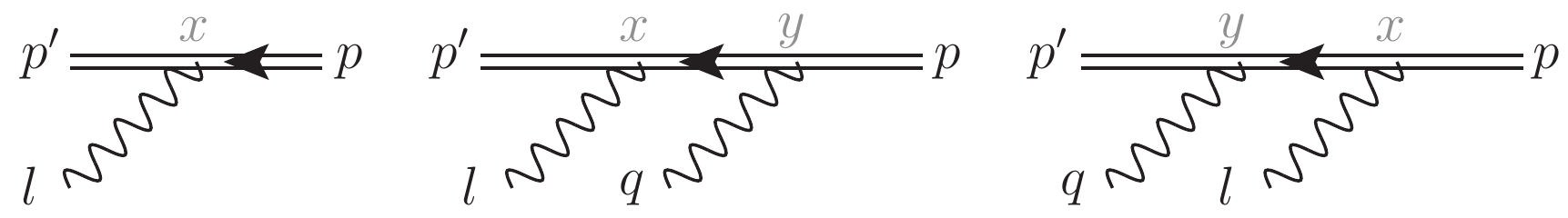}
\caption{\label{FIG:EN} The first diagram is NLC at tree level. Adding a laser-degenerate photon emission yields the second and third diagrams (`double' NLC at tree level,  {for investigations of which see~\cite{Seipt:2012tn,Mackenroth:2012rb,King:2014wfa,Dinu:2018efz}).}}
\end{figure}

We consider nonlinear Compton scattering (NLC), the emission of a photon of momentum $\ell_\mu$ and polarisation $\varepsilon_\ell^\mu$, from an electron of initial momentum $p_\mu$ scattering off the plane wave to momentum $p'_\mu$. Beginning at tree level, the scattering amplitude is, see also Fig.~\ref{FIG:EN},
\be\label{CALNDEF}
	\mathcal{N} := -ie\int\!\ud^4 x\, \bar\psi_{p'} (x) e^{i \ell\cdot x} \slashed{\varepsilon}_\ell \psi_p(x)\;,
\ee
the detailed evaluation of which is well-covered in the literature~\cite{Ritus1985a,Harvey:2009ry,Boca:2009zz,DiPiazza:2011tq,Seipt:2017ckc,Adamo:2020qru}; the only detail we will need here is that the $\{x^\LCperp,x^\LCm\}$ integrals in any part of an amplitude on a plane wave background can be performed to yield momentum-conserving delta functions in \textit{three} directions. For NLC this means
\be\label{delta}
	\mathcal{N} \propto \delta^3_{\LCperp,\LCm}(p' + \ell - p) \;.
\ee
We next consider the same process but with an additional, unobservable, photon emission (almost) degenerate with the plane wave, i.e.~{with photon momentum $q_\mu =q_\LCp n_\mu$.} The scattering amplitude is
\be
	\mathcal{M}_2 := (-ie)^2\int\!\ud^4 x\,\ud^4 y\,  \bar\psi_{p'} (x) e^{i \ell\cdot x} \slashed{\varepsilon}_\ell S(x,y) \slashed{\varepsilon}_q e^{i q\cdot y} \psi_p(y)  +\bar\psi_{p'}(y) \slashed{\varepsilon}_q e^{i q\cdot y}S(y,x)e^{i \ell\cdot x} \slashed{\varepsilon}_\ell \psi_p(x)  \;,
\ee
in which the second term is the photon exchange diagram, and where $\varepsilon^{\mu}_{\ell}$, $\varepsilon^\mu_q$, are the polarisation vectors for the photons of momenta $\ell_\mu$ and $q_\mu$ respectively. Applying (\ref{FromLeft}) and (\ref{FromRight}) we find
\be
\begin{split}\label{DNLC}
	\mathcal{M}_2 = (-ie)^2\int\!\ud^4 x\, &\bar\psi_{p'}(x) e^{i\ell.x} \bigg[ \frac{i \slashed{\varepsilon}_q\slashed{n} \slashed{\varepsilon}_\ell}{2n\cdot p'}  + \frac{i \slashed{\varepsilon}_\ell\slashed{n} \slashed{\varepsilon}_q}{2n\cdot p} \bigg] {e^{i q\cdot x}}\psi_p(x) \\
	&+\bar\psi_{p'}(x) e^{i \ell\cdot x}   \slashed{\varepsilon}_\ell  \psi_p(x)  \bigg(  \int\limits^{\infty}_{x^\LCp}\!\ud y^\LCp\, \frac{\varepsilon_q\cdot \pi'(y^\LCp)}{n\cdot p'}e^{
	iq_\LCp y^\LCp} + \int\limits_{-\infty}^{x^\LCp}\!\ud y^\LCp\, \frac{\varepsilon_q\cdot \pi(y^\LCp)}{n\cdot p}e^{
	iq_\LCp y^\LCp}\bigg)\;.
\end{split}
\ee
Recall that our interest is in the behaviour of this, and related, processes at high intensity, i.e.~large $a_0$. We will therefore keep track (in addition to possible soft divergences) of the leading-order $a_0$ dependence introduced by collinear effects, much as one might only keep track of leading logs in an IR calculation.

Observe that in the second line of (\ref{DNLC}) , the term \textit{outside} the large round brackets is just the integrand of the NLC scattering amplitude $\mathcal{N}$, as in (\ref{CALNDEF}). Multiplying this, inside the brackets, are the two scalar terms from (\ref{FromLeft}) and (\ref{FromRight}). Now, we {can choose to work in} lightfront gauge such that $n\cdot\varepsilon=0$ for photon polarisation vectors $\varepsilon$; hence we can set, directly from (\ref{Lorentz}), $\varepsilon_q \cdot \pi = \varepsilon_q \cdot (p-a)$, so that we have $\varepsilon_q \cdot \pi \sim -\varepsilon_q \cdot a \sim a_0$ for $a_0$ large; for a moment we retain also the $\varepsilon_q \cdot p$ term, as this carries the IR divergence in (\ref{vaclimit}). Compared to these terms, those in the first line of (\ref{DNLC}) are subleading in both the IR and in $a_0$, so we drop them.  Here we would seem to be stuck -- there is no further simplification we can make in general. In particular, the $x^\LCp$-dependence of the integrals stops us from factorising {collinear contributions out of the amplitude, in contrast to soft effects~\cite{Ilderton:2012qe}.}

\subsubsection{Factorisation and exponentiation}
There is however a subset of observables for which collinear corrections factorise. Consider that part of the photon spectrum for which $s:=n\cdot l /n \cdot p \ll 1$ for the emitted photon. As discussed in the main text, this does not mean the emission is itself collinear with the laser, even though laser photons have {$s = n\cdot q / n\cdot p = q_\LCm / p_\LCm = 0$}. In this regime, we may replace, in the expressions above,
\be\label{antagandet}
	\frac{1}{n\cdot p'} = \frac{1}{n\cdot p} \bigg( 1 + \frac{n\cdot l}{n \cdot p}  + \ldots \bigg) \simeq 	\frac{1}{n\cdot p} \;,
\ee
in which we have used the momentum conservation rule in (\ref{delta}), which continues to hold because the emissions we are adding are collinear to the laser.   As in the main text, we use a prime on amplitudes to indicate that we restrict to that part of the spectrum for which~(\ref{antagandet}) applies up to some acceptable error. With this assumption, the \textit{field-dependent} parts of the scalar factors in (\ref{DNLC}) combine into a single integral {which is $x$-independent and computes} the Fourier transform of the background. We then have
\be\label{9}
	\mathcal{M}'_2 \sim -ie\bigg( i \frac{\varepsilon_q \cdot p'}{q \cdot p'} - i\frac{\varepsilon_q \cdot p}{q \cdot p}  - \frac{\varepsilon_q\cdot \tilde{a}(q_\LCp)}{n\cdot p}\bigg) \mathcal{N}' \;.
\ee
Hence, in the given part of the spectrum, the effect of adding a collinear emission is simply to multiply the $S$-matrix element by a scalar factor. Mod-squaring, summing over polarisations $\varepsilon^\mu_q$ and integrating over $q_\mu$ using the {measure~(\ref{mott1}) or~(\ref{mott2}) yields}
\be\label{soft-col}
	|\mathcal{M}_2'|^2 \simeq \bigg(-\frac{\alpha \theta_0^2}{4\pi}\int\limits_{\mu}^{\omega_0}\! \ud q_\LCp \frac{1}{q_\LCp} \bigg( \frac{p'}{n \cdot p'} - \frac{ p}{n \cdot p}\bigg)^2  + q_\LCp \frac{ \tilde{a}(q_\LCp)\cdot \tilde{a}^*(q_\LCp)}{(n\cdot p)^2}  + \ldots \bigg) |\mathcal{N}'|^2 \;.
\ee
The ellipses denote terms which are IR finite and of lower order in $a_0$.

\begin{figure}[t!]
\includegraphics[width=0.45\textwidth]{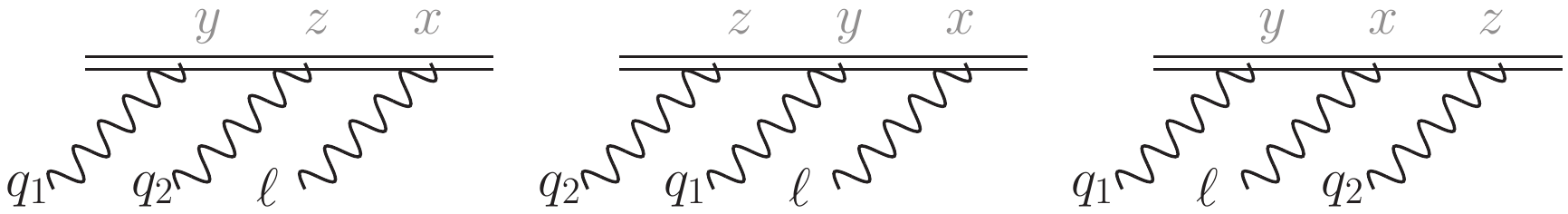}
\includegraphics[width=0.45\textwidth]{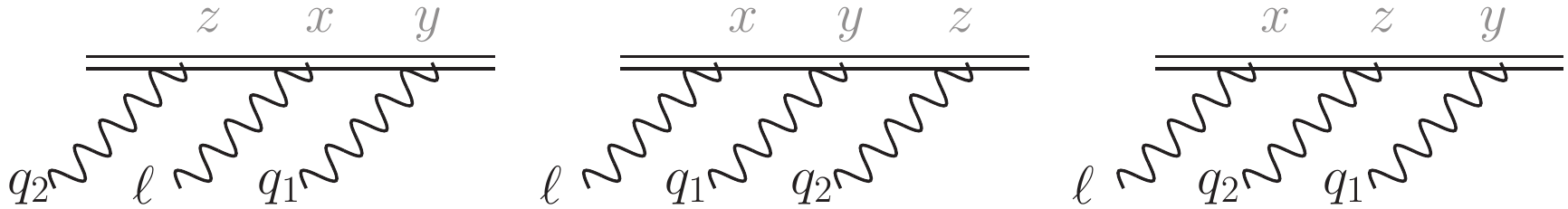}
\caption{\label{FIG:DNLC} The six diagrams contributing to `triple' NLC, in which two emissions are degenerate with the laser.}
\end{figure}

{Now consider the analogous processes with two or more degenerate emissions.} These emissions again generate lightfront time-ordered integrals, as appear in (\ref{DNLC}). In the primed regime, these again combine to give $x^\LCp$-independent factors, as we will demonstrate here for the case of two collinear emissions\footnote{{By momentum conservation, for many \textit{almost} collinear emissions it becomes a better and better approximation to stay in the `primed' regime, i.e.~taking $n\cdot l/n\cdot p$ small, as more energy is carried away by the collinear photons.}}. Accounting for photon exchange, there are six Feynman diagrams, see Fig.~\ref{FIG:DNLC}. Let the first collinear photon $q_1$ always be attached to vertex $y$, and $q_2$ {be attached} to vertex $z$. The non-degenerate photon with momentum $\ell$ is always at vertex~$x$. In each diagram, the leading-order integrands are the same, in the primed regime, as they were above, except each features integrals over different regions of $x$-$y$-$z$ space. Let us write $(xy)$ as shorthand for the step function $\Theta(x^\LCp - y^\LCp)$, then the six contributions can be summarised as, matching the order of the diagrams in Fig.~\ref{FIG:DNLC},
\be
	(yz)(zx) + (zy)(yx) + (yx)(xz) + (zx)(xy)  + (xy)(yz) + (xz)(zy) \;.
\ee
Observe that we can write, for example, the first term as $(yz)(zx) = (yz)(zx) \overline{(yx)}$ in which the bar indicates that the new step-function is implied by the first two. Using this trick, consider the sum of the first three diagrams:
\be\begin{split}
	(yz)(zx)\overline{(yx)} + (zy)(yx)\overline{(zx)} + (yx)(xz) &= (zx)(yx)\big[(yz)+(zy)\big] + (yx)(xz)  \\
	&= (zx)(yx)+  (yx)(xz) \\ & = (yx) \;.
\end{split}
\ee
Similarly, the second three diagrams sum to $(xy)$, so that in total one integrates over `1' meaning the whole $y$-$z$ plane. Thus, to the level of approximation considered, the collinear contribution {becomes $x$-independent and again factorises}. Squaring up, we obtain the same contribution as for one emission, but squared, and with an additional $1/2!$ symmetry factor for two collinear photons -- in other words the second term of the exponential series. The extension to higher orders is then clear.

Performing the incoherent sum over all possible numbers of degenerate emissions therefore gives an exponential correction to the exclusive NLC probability. The leading order behaviour, in intensity, of the inclusive NLC probability with respect to laser-collinear emissions is

\be\label{M-INC-TREE}
	|\mathcal{M}'|^2_\text{inc.} \sim \exp\bigg[-\frac{\alpha \theta_0^2}{4\pi}\int\limits_{\mu}^{E}\!\frac{ \ud q_\LCp }{q_\LCp} \bigg( \frac{p'}{n \cdot p'} - \frac{ p}{n \cdot p}\bigg)^2\bigg]\,\exp \bigg[  \frac{\alpha \theta_0^2}{4\pi}\int\limits_{\mu}^{E}\! \ud q_\LCp\,  q_\LCp \frac{ |\tilde{\sf a}(q_\LCp)|^2}{(n\cdot p)^2} \bigg] |\mathcal{N}'|^2 \;.
\ee
The first exponential is simply a piece of the usual soft IR divergence~\cite{Yennie:1961ad,Weinberg:1965nx,Laenen:2008gt,Gardi:2010rn}, since our degenerate integration region includes a (spherical angular) portion of soft parameter space. {However we already know that this will cancel against loop corrections (nor {is it} explicitly field-dependent).} We therefore drop the {IR divergent} contribution from here on. Our interest is in the second exponential of (\ref{M-INC-TREE}). This is, like the soft correction, positive, but explicitly increasing with intensity $a_0$. We note that the exponent depends on intensity $a_0$ and electron lightfront energy $n\cdot p$ individually, and \textit{not} {on their product} $\chi := a_0 \omega_0 n.p/m^2 $, which is the only parameter on which (locally) constant crossed-field results depend. ({A dependence on intensity and energy separately is the behaviour one expects of fields more general than CCFs~\cite{Dinu:2013gaa,Podszus:2018hnz,Ilderton:2019kqp}, and indeed is the reason why the high-energy and high-intensity limits, which both yield high $\chi$, do not commute beyond CCFs~\cite{Podszus:2018hnz,Ilderton:2019kqp}.}) This shows that LCFA calculations cannot capture the effects we consider.  Observe that, at large enough $a_0$, the exponential growth of the real emission correction violates unitarity; this growth must therefore be compensated for by loop corrections, to which we {turn in the next Section. Before moving on, though, we comment briefly on the form of similar collinear corrections in another processes, namely nonlinear Breit-Wheeler pair production, the amplitude for which is obtained by crossing symmetry from that of NLC. 

In nonlinear Breit-Wheeler, adding a laser-collinear emission onto either the outgoing positron or electron line gives a lightfront time integral like that in (\ref{FromRight}), from the vertex position (denoted $x$ above, e.g.~in (\ref{DNLC})) to infinity, rather than one as in (\ref{FromLeft}) and one as in (\ref{FromRight}).  This suggests that we would not be able to factorise out a spatially independent factor from the collinear insertion (at least under the same hypothesis as in the text), because we could not combine the two different integration regions to remove the $x$-dependence. Consequently, a much more involved calculation would be needed to analyse collinear corrections for nonlinear Breit-Wheeler. We speculate that this may be related to a behaviour seen in~\cite{Mironov:2020gbi} for higher loop diagrams in constant crossed fields. There, two different intensity-dependencies were seen to arise in resummed bubble-chain diagrams, attributed to different possible cuts of the loop amplitudes into either photon-producing processes (like NLC) or pair-producing processes (like nonlinear Breit-Wheeler). This rather intriguing story will be addressed elsewhere. Here we return to NLC, and proceed to consider loop corrections.  }

\section{virtual collinear corrections to NLC}\label{APPB}
\setcounter{subsubsection}{0}
As the inclusive probability above contains all orders in $\alpha$, we must for consistency add all order loop corrections. It is not possible to do this exactly, but we will show in this section that we can include the laser-collinear parts of the loops and see how they impact high-intensity {scaling. For this we will need the photon propagator. We} saw above that collinear emissions generate lightfront time-ordered terms, and that the leading-$a_0$ dependence of our amplitudes was easily identified using lightfront {gauge. In this gauge the the photon propagator is}
\be\label{GDEF}
\begin{split}
	G_{\mu\nu}(y-z) &= -i \int\!\frac{\ud^4 q}{(2\pi)^4} \frac{e^{-iq\cdot(y-z)}}{{q^2+i\epsilon} } \bigg(g_{\mu\nu}  - \frac{n_\mu q_\nu + q_\mu n_\nu}{n\cdot q}\bigg) \;.
\end{split}
\ee
It is convenient to immediately evaluate the $q_\LCp$ integral in $G_{\mu\nu}$ using the residue theorem, which yields
\be
\begin{split}\label{G-int-done}
	G_{\mu\nu}(y-z)
	  &= -\int\!\frac{\ud^2q_\LCperp\ud q_\LCm}{(2\pi)^32|q_\LCm |} \, e^{-iq\cdot(y-z)}  \Theta\big(q_\LCm(y-z)^\LCp\big)  L_{\mu\nu}{(q)} -\int\!\frac{\ud q_\LCm}{2\pi}  e^{-iq_\LCm(y-z)^\LCm} \frac{i\delta^{\LCp,\LCperp}(y-z)}{2 q^2_\LCm} n_\mu n_\nu \;.
\end{split}
\ee
in which $q_\LCp$ is now determined by the on-shell condition, and the `transverse projector' $L_{\mu\nu}(q)$ is the same tensor as in (\ref{GDEF}), but for $q_\mu$ on-shell. While (\ref{G-int-done}) is a commonly used result, it is worth highlighting the features relevant to our {investigation}. Consider the first term in (\ref{G-int-done}). We note that i) $L_{\mu\nu}$ is orthogonal to both $q_\mu$ and $n_\mu$, as were our polarisation tensors for real photons, ii) the step function will introduce into amplitudes an additional ligtfront-time ordered dependence which, recall from (\ref{FromLeft}) and (\ref{FromRight}), is the essential structure introduced by real collinear emissions, iii) the integration measure is almost the on-shell measure, as appeared for real emissions. It is for these reasons that lightfront gauge is natural for our problem~\cite{Bakker:2013cea}.  The second term in (\ref{G-int-done}) is the instantaneous propagator, which we will see below {yields terms subleading in $a_0$}. IR cutoffs are in place just as in the tree-level calculation.

We begin, as we did for real emissions, with the leading order contributions in $\alpha$, before considering the generalisation to {all orders. The order $\alpha$ contribution to account for is}  the interference term between the tree-level and one-loop NLC amplitudes which arises when the whole amplitude is mod-squared. There are three contributing one-loop diagrams\footnote{The remaining one-loop contribution to NLC contains the photon polarisation tensor, but no photon loop, and hence does not contribute to our discussion of {laser-collinear} corrections. We note also that the polarisation tensor inserted onto a laser-collinear photon will reduce to the {free loop~\cite{Ilderton:2016qpj}}, and hence be intensity independent.}, shown in Fig.~\ref{FIG:LOOPS}. These are, respectively, the vertex correction~\cite{Morozov:1981pw,Gusynin1999,DiPiazza:2020kze} and two self-energies~\cite{Ritus1985a,Meuren:2015lbz}.   UV divergences are removed by subtracting the free-field contributions as usual, but we are only interested in $a_0$-dependent terms, which are UV-finite. We consider each in turn.

\subsubsection{The vertex correction.}
The expression to evaluate is
\be\label{vertex}
	\mathcal{N}_\text{vertex} := (-ie)^3 \int\!\ud^4 x\,\ud^4 y\,\ud^4 z \, \bar{\psi}_{p'}(y) \gamma^\mu S(y,x) \slashed{\varepsilon} e^{i\ell\cdot x} {S(x,z)} \gamma^\nu \psi_p(z) G_{\mu\nu}(y-z) \;.
\ee
We would like to evaluate the $y$ and $z$ integrals in (\ref{vertex}), but taking account only of laser-collinear effects, or in other words isolating the contribution for which the collinear results (\ref{FromLeft}) and (\ref{FromRight}) can be applied.  Because we can take $q_\mu$ to be on-shell in the integral representation~(\ref{G-int-done}) of $G_{\mu\nu}$, we know from the above that \textit{some} appropriate integration range exists as it covers at least the region appropriate to real emissions. We specify the {full region in the main text, but the details are not needed at this stage: we} need only know that in this region we can apply (\ref{FromLeft})--(\ref{FromRight}).

We begin with the instantaneous term in the propagator. This contributes, highlighting only the relevant terms, and underlining those coming from the propagator,
\be\label{no-inst}
	\mathcal{N}_\text{vertex} \sim 	\ldots \int\!\ud^4 y\, \bar{\psi}_{p'}(y) \gamma^\mu \dashuline{e^{-iq \cdot y}}  S(y,x) \ldots \dashuline{n_\mu n_\nu} \int\!\ud^4 z \, {S(x,z)} \gamma^\nu \dashuline{e^{iq \cdot z}}  \psi_p(z)  \dashuline{\delta(y^\LCp-z^\LCp)} \;.
\ee
Using (\ref{FromLeft})--(\ref{FromRight}), the generated factors of $\pi_\mu$ will be contracted with the highlighted factors of $n_\mu$ from the propagator. However $n\cdot\pi = n\cdot p$, from (\ref{Lorentz}), which is independent of $a_0$. Hence the instantaneous propagator does not contribute to the \textit{leading-order} intensity dependence. (The highlighted delta function in (\ref{no-inst}) does not affect this argument, because the $y^\LCp$ and $z^\LCp$ integrals do not need to be, and are not, evaluated for (\ref{FromLeft}) or (\ref{FromRight}) to hold.) We turn to the $L_{\mu\nu}$ term in the propagator. Again writing out only the important structures {which will generate terms leading in intensity}, and underlining contributions from the propagator, we find
\be\label{TRE-THETA}
	\mathcal{N}_\text{vertex} \sim \ldots \int\limits_{x^\LCp}\!\ud y^\LCp  \dashuline{e^{-iq\cdot y}} \frac{\pi^{\prime\mu}(y^\LCp)}{n\cdot p'}  \ldots \dashuline{L_{\mu\nu}(q)} \ldots 	\int\limits^{x^\LCp}\!\ud z^\LCp \frac{\pi^\nu(z^\LCp)}{n\cdot p} \dashuline{e^{iq\cdot z}}\,   \dashuline{\Theta[q_\LCm(y^\LCp-z^\LCp)] }  \;.
\ee
Observe that the time-ordering in the integrals simplifies the {Heaviside} theta-function in the propagator to just $\Theta(q_\LCm)$, which turns the measure on $q^\mu$ in (\ref{G-int-done}) into the on-shell measure {proper, see (\ref{os-mott}).} Evaluating $\pi' \cdot L(q) \cdot \pi$, the leading intensity dependence is $a(y)\cdot a(z)$. Thus we find the {laser-collinear} part of the vertex correction is, to leading order in intensity,
\be\label{BIDRAG1}
	\mathcal{N}_\text{vertex} \sim e^2 \int\!\ud q_\text{o.s.} \int\!\ud^4x\, \bigg[-ie\, \bar{\psi}_{p'}(x) \slashed{\varepsilon} e^{i\ell\cdot x} \psi_p(x) \bigg] \int\limits_{x^\LCp}^\infty\!\ud y^\LCp \!\!\int\limits_{-\infty}^{x^\LCp}
\!\ud z^\LCp {e^{-iq_\LCp  y^\LCp}}\frac{a(y)\cdot a(z)}{n\cdot p' n\cdot p}e^{i q_\LCp z^\LCp} \;.
\ee
This expression shows that the laser-collinear virtual contribution has the effect of multiplying the tree-level NLC  \textit{integrand}, shown in large square brackets, by a lightfront-time dependent scalar factor. As such there is no factorisation, although this should not surprise us given the tree-level results.

\subsubsection{The self-energies.}
The calculation of the self-energy diagrams proceeds as for the vertex correction. Subtleties concerning self-energy contributions on external legs, and renormalisation, are discussed in~\cite{Weinberg:1995mt}. However, we are interested in the UV finite, $a_0$-dependent terms, which contain the physics of e.g.~radiative spin flip~\cite{Ilderton:2020gno}, and must be included. We use a uniform notation such that the `hard' NLC vertex is always at $x$, the `left hand' end of the loop is at $y$, and so comes with a factor $e^{-iq\cdot y}$, while the right hand end of the loop is always at $z$, and comes with a factor $e^{iq\cdot z}$, see Fig.~\ref{FIG:LOOPS}.  The time-orderings and $\Theta$-functions appearing again conspire so as to extract a particular contribution from the propagator. The self-energy corrections are precisely as in (\ref{BIDRAG1}), except that the lightfront time integrals are replaced by 
\be\label{BIDRAG23}
	\int\limits_{x^\LCp}^{\infty}\!\ud z^\LCp \!\!\int\limits^{\infty}_{z^\LCp} \!\ud y^\LCp {e^{-iq_\LCp y^\LCp}}\frac{a(y)\cdot a(z)}{(n\cdot p')^2}e^{i q_\LCp z^\LCp} + 	\int\limits^{x^\LCp}_{-\infty}\!\ud y^\LCp \!\!\int\limits_{-\infty}^{y^\LCp} \!\ud z^\LCp e^{-iq_\LCp y^\LCp}\frac{a(y)\cdot a(z)}{(n\cdot p)^2}{e^{i q_\LCp z^\LCp}} \;.
\ee
%
\subsubsection{Factorisation and exponentiation}
Each of the one-loop contributions in (\ref{BIDRAG1}) and (\ref{BIDRAG23}) covers a different region in configuration ($x$--$y$--$z$) space. This dependence arises through lightfront-time ordering at both the vertices and from the photon propagator. It is, just as at tree level, the residual $x$--dependence of these time-orderings which prevents us from factorising out the laser-collinear contributions. Let us therefore make the same assumption as {for emissions}; we restrict attention to that part of the emitted photon spectrum with $n\cdot l \ll n\cdot p$ which, recall, allows us to replace factors of $n\cdot p'$ with $n\cdot p$. In this case the three $(y,z)$--integrands in (\ref{BIDRAG1}) and (\ref{BIDRAG23}) become equal. Writing the three regions of integration in the step-function notation introduced above, the sum of (\ref{BIDRAG1}) and (\ref{BIDRAG23}) may be represented as
\be\begin{split}
	(yx)(xz) + (zx)(yz) + (xy)(yz) &= (yx)(xz)\overline{(yz)} + (zx)(yz) + (xy)(yz)\overline{(xz)} \\
	&= (yz)(xz)\big[ (yx) + (xy) \big] +  (zx)(yz) \\
	&=  (yz) \big[(xz) +  (zx)\big] \\ &= (yz) \;,
\end{split}
\ee
which is $x$-independent. Thus, the sum of all one-loop contributions factorises, and we can write
\be
	\mathcal{N}^\prime_\text{1-loop} \sim \bigg[e^2\int\!\ud q_\text{o.s.} \int\limits_{-\infty}^\infty\!\ud y^\LCp \!\!\int\limits_{-\infty}^{{y^\LCp}}
\!\ud z^\LCp\, {e^{-iq_\LCp y^\LCp}}\frac{a(y)\cdot a(z)}{(n\cdot p)^2}{e^{i q_\LCp z^\LCp}} \bigg] \mathcal{N}^\prime \;.
\ee
The remaining time-ordering can be undone to again yield Fourier transforms of the field, {at the cost of introducing a factor of $1/2$. This factor is cancelled when we mod-square the sum of the tree-level and one loop contributions, to find}
\be
	|\mathcal{N}' + \mathcal{N}^\prime_\text{1-loop}|^2 \simeq \bigg[1 - e^2\int\!\ud q_\text{o.s.}  \frac{|\tilde{\mathsf{a}}(q_\LCp)|^2}{(n\cdot p)^2} \bigg] |\mathcal{N}^\prime|^2 +\mathcal{O}(e^4) \;.
\ee
which is of precisely the same form as the contributions from real emissions, the only difference being in the (here unspecified) integration region. 

For the extension to higher orders, we note that {the integration regions in the three one-loop amplitudes combine, in the primed regime}, to cover the whole $y$-$z$ plane, forming an $x$-independent factor (factorisation). By comparison with the free theory, and through the explicit examples above, one sees that the addition and combination of the different loop contributions on the external fermion legs corresponds in Fourier space to the same combinatorics of combining soft legs into {loops (as is explicit in the appearance of the soft factors for real emissions in (\ref{soft-col})). We are only adding collinear photons to the fermion legs in our diagram, the effect of which is to multiply by scalar factors when we sum over all permutations, hence all fermion legs remain `effectively' external in our NLC-degenerate processes, and at higher orders the counting of diagrams and factors goes through as above. Hence the virtual contributions will also factorise at higher orders, just as they did for real emissions, and exponentiate, as they must to maintain unitarity at high intensity. 

Indeed there is a sense in which the exponential form of the laser-collinear contributions should not be surprising; the Volkov solutions themselves can be viewed simply as encoding the summation over all possible \textit{coherent} emissions and absorptions of mutually collinear laser photons~\cite{Frantz,Gavrilov:1990qa,Ilderton:2017xbj}, and are themselves exponential functions of intensity.}

\section{phenomenology}\label{APP:PHEN}
For a phenomenological estimate, we take a simple, circularly polarised Gaussian pulse; the two-component transverse potential is 
\be
	{{\sf a}}(x) = m a_0\, 2^{-\frac{2x^2}{\tau^2}} \big(\cos \omega_0 x ,\sin \omega_0 x \big) \;,
\ee
with normalisation chosen such that $a_0$ is the peak intensity and $\tau$ is the full-width half-maximum (intensity) pulse duration. We consider the case of a long, almost flat-top pulse, which means $\tau$ large, or, in Fourier space, narrow bandwidth. In this case we may approximately write the Fourier transform of the field as
\be
	|\tilde{{\sf a}}(q)|^2 \sim \frac{\pi^{3/2} \tau}{\sqrt{\log 16}} \delta_\text{reg}(q-\omega_0)+  \ldots 
\ee
in which $\delta_\text{reg}$ is a properly normalised delta-function regulated by $\tau$, and the ellipses denote a second spectral delta which does not contribute to our results. We can now easily evaluate the exponential factor in (\ref{everything2}).  We find
\be
\frac{-\alpha\theta_0^2\pi^{3/2}}{8\pi\sqrt{\log(16)}} (\omega_0\tau)^2 \frac{a_0^2m^2}{(n\cdot p)^2} 
 \simeq -5\cdot 10^{-5} \frac{a_0^2T}{\gamma^2(1+\beta)^2} \;,
 \ee
in which {the integral} over the regularised delta yields a factor of $1/2$, and for the final `engineering formula' we have assumed a head-on collision and set $\theta=\tan^{-1}0.15\simeq 0.15$ for the opening angle of a typical optical laser with $\omega=1.55$~eV~\cite{ELI}, and $T$ is the FWHM pulse duration measured in femtoseconds. To maximise the impact of collinear corrections we need intense, long, pulses, and low energy electrons; for large $\gamma$, the collinear contribution is very strongly suppressed. Indeed, it is clear that, for any experiment which could be performed in the very near future, the corrections implied by (\ref{everything2}) are negligible, but this is not the regime of interest.

\end{document}